# A model for enhanced fusion reaction in a solid matrix of metal deuterides

K. P. Sinha[a] and A. Meulenberg[b]

[a] Department of Physics, IISc, Bangalore 560012, India (kpsinha@gmail.com)
[b] HiPi Consulting, Frederick, MD, USA (mules333@gmail.com)

## Abstract

Our study shows that the cross-section for fusion improves considerably if d-d pairs are located in linear (one-dimensional) chainlets or line defects. Such non-equilibrium defects can exist only in a solid matrix. Further, solids harbor lattice vibrational modes (quanta, phonons) whose longitudinal-optical modes interact strongly with electrons and ions. One such interaction, resulting in potential inversion, causes localization of electron pairs on deuterons. Thus, we have attraction of $D^+ - D^-$ pairs and strong screening of the nuclear repulsion due to these local electron pairs (local charged bosons: acronym, lochons). This attraction and strong coupling permits low-energy deuterons to approach close enough to alter the standard equations used to define nuclear-interaction cross-sections. These altered equations not only predict that low-energy-nuclear reactions (LENR) of $D^+ - D^-$ (and $H^+ - H^-$) pairs are possible, they predict that they are probable.

## Introduction

The central idea of this paper is that the solid matrix in which LENR takes place provides conditions such as: (1) confinement of deuterons (or deuterium atoms) in linear chains or in line defects, (2) dynamical solid-state modes (phonons), which can store and exchange energy, and (3) the strong interaction between appropriate phonon modes (particularly longitudinal-optical modes) with electrons and ions. The resultant resonant $D^+$-$D^-$ pairs in this environment permit attractive forces and/or strongly-screened repulsive forces, rather than the normally-expected strong-repulsive Coulomb forces between positively-charged nuclei. These options are not available in a gaseous plasma, or in materials that lack (at least) short-range order.

It is a goal of this paper to provide an understandable, standard-physics basis (under special conditions) for the extensive body of results presently available from LENR.[1]

## Model

Let us consider a linear chain of deuterons surrounded by an equal number of electrons. The Hamiltonian for such a system is [2]

$$H = H_e + H_L + H_{eL} + A, \qquad (1)$$

where the electron contribution is

$$H_e = \Sigma_\sigma E_m C^+_{m\sigma} C_{m\sigma} + \Sigma\, t_{mn}\ (C^+_{m\sigma} C_{m\sigma} + h.c). \tag{2}$$

Here the $C^+_{m\sigma}$ ($C_{m\sigma}$) denote the electron creation (annihilation) operators in the Wanneir state $|m\sigma\rangle$, at site "m" with spin $\sigma$. $E_m$ is the onsite single-particle energy of the electron and $t_{mn} = -|t|$ is the nearest-neighbor hopping integral. The lattice Hamiltonian is

$$H_L = \hbar\omega_D \Sigma_m (d^+_m d_m + \tfrac{1}{2}), \tag{3}$$

where $\omega_D$ is the vibration frequency of the deuterium atom D (taken as an Einstein oscillator) and with $d^+_m$ ($d_m$) denoting the phonon creation (annihilation) operators.

The interaction of electrons with the above phonon modes is described by

$$H_{eL} = g\hbar\omega_D \Sigma_\sigma C^+_{m\sigma} C_{m\sigma} (d^+_m + d_m), \tag{4}$$

where g is a dimensionless coupling constant. The last term in Equation 1, A, is a constant negative energy due to negative space charge in the channel. Note that in a low-dimension (one or two) structure, the potential energy between two deuterium atoms is much deeper and negative, relative to that of atoms in a 3-D lattice.[3] A suitable unitary transformation[4,5] leads to a displaced harmonic oscillator $[d'_m \to (d_m + \delta)]$ and, in the transformed total Hamiltonian, the on-site single-electron energy $E^*_m = E_m - E_d$, with $E_d = g^2 \hbar\omega_D$; the hopping integral (in Equation 2)

$$t^*_{mn} = t_{mn} \exp(-g)^2; \tag{5}$$

and the electron effective mass[2]

$$m^* = m_e \exp[E_d/\hbar\omega_D] = m_e \exp[g^2]. \tag{6}$$

Even for a very conservative value of $g^2 = 1.6$, this will give $m^* = 5m_e$ (see below Equation 9).

Let us now consider the situation of two deuterons and two electrons in a chain. This introduces Coulomb repulsion ($U_e$) between two electrons about an atom at site "m" in the same orbital state $|m\rangle$, but having opposite spin. The displacement transformation

$$(C^+_m)^* = C^+_m \exp[-g (d_m - d^+_m)], \tag{7}$$

gives the effective Hamiltonian and the various parameters are obtained as

$$E^*_m = E_m - E_d; \qquad U_e^* = U_e - 2 E_d; \qquad \text{and } t^* = |t| \exp[-g^2] \tag{8}$$

For $U < 2E_d$, $U^*$ becomes negative. Thus, there is potential inversion for the 2 electrons in the singlet state and they will form a small on-site localized pair, a sort of composite boson (lochon)[2,6]. Under this condition, the $D^-$ state will be more stable (has lower energy) than the neutral atom D (though not necessarily more stable than the $D^+ = d$ state). This would lead to the existence of $D^+$-$D^-$ pairs. They would exist in the resonating state, $D^-$-$D^+ \leftrightarrows D^+$-$D^-$, further reducing their energy and inter-nuclear distance.

## Strong Screening

Bound electrons reduce the effective charge of nuclei. An occasional transfer of one such electron between two deuterium atoms forms a transient electron pair within a $D^+$-$D^-$ pair. At separations larger than the orbital radius of the electrons, this transfer changes a neutral relationship to an attractive one. At separations smaller than a fraction of the orbital radius of the electrons, it still gives a significant reduction in "effective" Coulomb repulsion between the nuclei. This effective potential will be represented by

$$U_d^* = ((e^*)^2/r) = (e^2/r)(1 - \exp[-a_s/\lambdabar_L]) \quad , \tag{9}$$

where $a_s$ is the strong-electron-screening length, $\lambdabar_L = (\hbar/m_L^* \upsilon_L)$ is the rationalized deBroglie wavelength[1] of the lochon, $m_L^*$ being the effective mass of the lochon, and $\upsilon_L$ its speed (as determined from its energy in the atomic and molecular potential wells of the two deuterons). This screening by lochons is a short-range effect and reduces the repulsive potential between reacting nuclei (deuterons here). Screening by itinerant electrons is weak in this range (relative to that of the bound electrons) and hence not considered here.[7]

The coupling into an optical-phonon mode, along with the attractive potential of the $D^+$-$D^-$ pair, briefly produces a nearly 1-D encounter that greatly increases the potential-well depth of this short-lived "molecule."

## Penetration Factor and Cross Section

Next, we discuss the fusion reaction of a screened d-d reaction in 1-D. For an incident particle of effective charge $e^*$, the penetration factor $P(l, E_a)$ decreases rapidly with its decreasing total energy, $E_a$, where $l$ is its orbital angular-momentum state. In this low-energy situation, particles in an $l = 0$ state contribute most.[8] We have:

$$P(0, E_a) = (V_o(R)/E_a)^{1/2} \exp\left[-2\int_R^{r_0}(|k^2 - (2M_d/\hbar^2)(e^*)^2/r|)^{1/2} dr\right]$$

$$= (V_o(R)/E_a)^{1/2} \exp\left[-2k\int_R^{r_0}(|1 - (r_o/r)|)^{1/2} dr\right] \quad , \tag{10}$$

where k is the wave vector of particle $a$, $M_d$ is the reduced mass of two deuterons, and with $V_o(R) = (e^{*2})/R$:

$$r_o = (e^*)^2 \, 2M_d/\hbar^2 k^2, \qquad R \ll r_o. \tag{11}$$

The integral requires careful treatment since, as $r \to 0$, it has a singular term. Hence, resorting to an asymptotic expansion (a better approximation for expression of the integral as $r \to 0$), we get,

$$P(0, E_a) = (V_o(R)/E_a)^{1/2} \exp[-(e^*)^2/\hbar \upsilon_r] \quad , \tag{12}$$

where $\upsilon_r$ is the relative velocity. The cross-section $\sigma(a,b)$ of this reaction is

$$\sigma(a,b) = (\text{constant} / E_a) \exp[-(e^*)^2/\hbar \upsilon_r]$$

$$= (\pi/k^2) \exp[-(e^2/\hbar \upsilon_r)(1 - \exp(-a_s/\lambdabar_L))] \quad , \tag{13}$$

where $k^2 = (2 M_d E_a / \hbar^2) = M_d^2 \upsilon_r^2 / \hbar^2$ and $M_d$, $E_a$, $a_s$, and $\lambdabar_L$ are as above. The critical difference between this development[9] and the prior work (standard model) is a factor of $2\pi$ in the exponent that exists in the regular solution and is gone here (valid at least for $r \Rightarrow R$).

The key values in the present model are those calculated for the deBroglie wavelengths for the lochon and the deuterons, as a function of d-d gap, and the value taken for $a_s$. This value is the screening provided by the bound electrons/lochon and is given in Ichimaru[7] (page 9) based on the ion-sphere model. Normally this is ½ the sum of the atomic-orbital radii for the charge state of the two atoms [$a_{ij} = (a_i + a_j)/2$]. In our model, $a_{ij} = 0.53$A for the D-D case and, since we can ignore the radius of the bare deuteron, $a_{ij}$ = ~ 0.3A for the $D^+$ - $D^-$ case. So, we assume a range of values between the initial value $a_s = a_{ij}/2\pi$ (the Bohr radius divided by $2\pi$, since we are using $\lambdabar_L$ in the equation) and that of the 1-D case (as described above and under the appropriate circumstances), which will reduce $a_s$ by up to an order of magnitude.

---

[1] Several of the references herein use $\lambdabar$, rather than $\lambda$ (the deBroglie wavelength), in their equations. We follow suit.

The value of $\lambdabar_L$ varies from ~$10^{-9}$ down to ~$10^{-10}$ cm, while the lochons accelerate between the deuterons and the Coulomb field grows as the gap shrinks. This large lochon size, relative to the nuclear-interaction distances, is a major limitation for strong screening. However, being bound, the electron/lochon screening of the deuterons increases with kinetic energy (i.e., as orbits shrink), rather than decreasing as is the case for free electrons. The 1-D nature of the problem affects the electron s-orbital orientation, in that the electron/lochon direction of motion is along the d-d axis, and therefore this localization and the velocity-induced shrinkage of $\lambdabar_L$ (along the d-d axis) aids in the screening.

**Reaction Rate**

The reaction rate (per cm$^3$ per second) for d-d- fusion with lochon screening is given by

$$R_{dd} = r^*_{dd} K_B T \lambdabar_L^2 N_d^2/\hbar \ [(e^2/\hbar \, \upsilon_r)(1 - \exp(-a_s/\lambdabar_L)] \\ \times \exp[-(e^2/\hbar \, \upsilon_r)(1 - \exp(-a_s/\lambdabar_L))] \quad ; \quad (14)$$

where $r^*_{dd} = \hbar^2/2 M_N e^2$ is the nuclear Bohr radius for a pair of deuterons; $M_N$ is the average mass per nucleon $\approx 1.66 \times 10^{-24}$ gm; $K_B$ is the Boltzmann constant; T is the temperature (with $K_B T \lambdabar_L^2$ having dimensions of energy x area); and $N_d$ is the concentration of deuterons per unit volume.

The model presented in the foregoing section is more appropriate for reaction on the surface or defect-plane in the lattice. The reaction rate is the number of effective collisions of deuterons per unit area per sec. To convert to this picture, set $K_B T \lambdabar_L^2 \Rightarrow K_B T \lambdabar_L$ = energy x length and set $N_d \Rightarrow N_S$ = number of deuterons per unit area (rather than per unit volume).

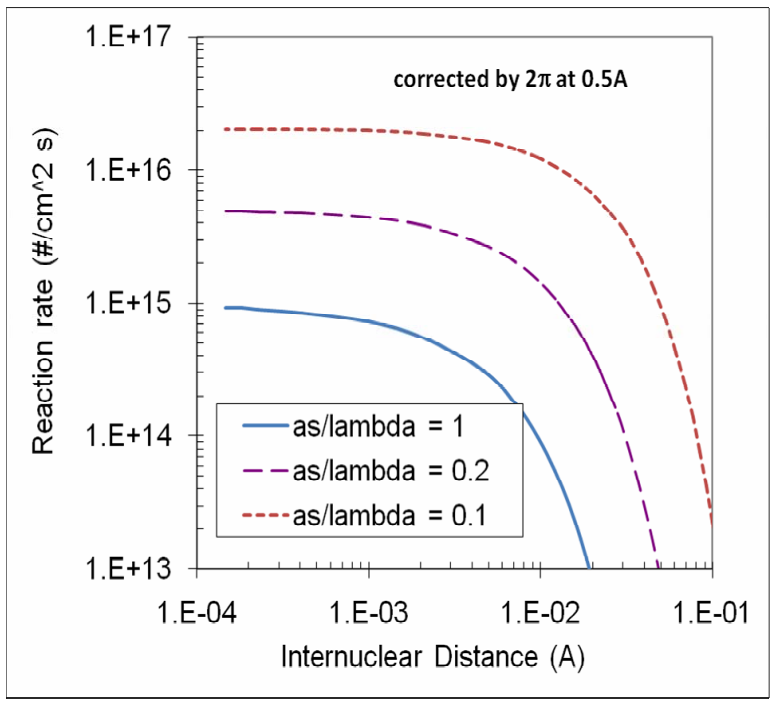

The results of Equation 14, modified for surfaces, are plotted in Figure 1, taking some acceptable values of the parameters involved. However, the figure plots the reaction rate assuming that 100% of the available lattice sites are actively involved. It is likely that only a percentage of the sites can be made to contribute to LENR. Three values of $a_s/\lambdabar_L$ (1, 0.2, and 0.1) have been selected for the lochon case.

*Figure 1   The $D^+$ - $D^-$ reaction rate (for a surface), as a function of $D^+$ - $D^-$ separation distance, for three ratios of $a_s/\lambdabar_L$ (1, 0.2, & 0.1).*

## Conclusion

In the foregoing sections, we have presented a model incorporating conditions in the condensed matter state that can facilitate fusion of deuterons aided by interaction of electrons with phonon modes of the system. The cross-section of the reaction improves considerably owing to the presence of d-d pairs in line defects and with strong screening provided by bound electron pairs (lochons). However, only a mechanism, such as $D^+$ and $D^-$ pairing can bring the deuterons close enough to permit a modified standard nuclear model to predict LENR.

Recent experiments by several workers, in which the material (e.g., powder or particles), is taken to be in the nanometer range, suggest that the creation of large surface area plays an important role.[10] These surfaces may provide the required active sites, in the 2-D geometry that can harbor lochons and $D^+ + D^-$ ion pairs. Our computed reaction rate is found to be $> 10^{14}$ per $cm^2$ per second (Figure 1) for two-dimensional surfaces, in agreement with the estimate of some workers.

The role of optical-phonon modes is important for their bringing the $D^+ + D^-$ pairs together, for coupling of ions to electrons, and as a source of resonant coupling to provide the required surface-mode excitation (surface plasmon or phonon) that can lead to enhanced-optical potentials. Recent work, on excitation of surface plasmon/polaritons with "tuned" lasers,[11,12] indicates the importance of this mechanism, where the induced-optical potential aids the fusion reaction by several orders of magnitude. The known presence of resonant $D^+ + D^-$ ion pairs in the solid state (coupled via optical phonons) greatly increases the d-d interaction cross-section by altering the shape of the Coulomb barrier to the extent of requiring a change in the equations normally used in nuclear physics.

## Acknowledgements

This work is supported in part by HiPi Consulting, New Market, MD, USA, by the Science for Humanity Trust, Bangalore, 560094, India, by the Science for Humanity Trust, Inc, Tucker, GA, USA, and by the Indian National Science Academy.